\documentclass{nature}
\usepackage{amssymb}
\usepackage{color}

\usepackage{graphicx}

\usepackage{bbold}
\usepackage{bm}
\usepackage{amsmath}
\usepackage{amsfonts}
\usepackage{amssymb}
\usepackage{braket}
\usepackage{units}

\usepackage{lineno}

\usepackage{csquotes}
\usepackage{fancyhdr}

\usepackage{graphicx}
\makeatletter
\let\saved@includegraphics\includegraphics
\AtBeginDocument{\let\includegraphics\saved@includegraphics}
\renewenvironment*{figure}{\@float{figure}}{\end@float}
\makeatother

\title{Competing magnetic orders in a bilayer Hubbard model with ultracold atoms}

\author{Marcell Gall$^\ast$, Nicola Wurz$^\ast$, Jens Samland, Chun Fai Chan, Michael K\"ohl}

\begin{document}
	
	\maketitle

\begin{affiliations}
	\item Physikalisches Institut, University of Bonn, Wegelerstra{\ss}e 8, 53115 Bonn, Germany
	\item $^\ast$ These authors contributed equally to this work.
\end{affiliations}

{\bf Fermionic atoms in optical lattices have served as a compelling model system to study and emulate the physics of strongly-correlated matter. Driven by the advances of high-resolution microscopy, the recent focus of research has been on two-dimensional systems \cite{Greif2015,Cheuk2016,Cocchi2016} in which several quantum phases, such as anti-ferromagnetic Mott insulators for repulsive interactions \cite{Cheuk2016b,Parsons2016,Drewes2017,mazurenko2017cold} and charge-density waves for attractive interactions  \cite{mitra2018quantum} have been observed. However, the aspired emulations of real materials, such as bilayer graphene, have to take into account that their lattice structure composes of coupled layers and therefore is not strictly two-dimensional. In this work, we realize a bilayer Fermi-Hubbard model using ultracold atoms in an optical lattice and demonstrate that the interlayer coupling controls a crossover between a planar anti-ferromagnetically ordered Mott insulator and a band insulator of spin-singlets along the bonds between the layers. Our work will enable the exploration of further fascinating properties of coupled-layer Hubbard models, such as theoretically predicted superconducting pairing mechanisms \cite{scalettar1994magnetic,maier2011pair}.}

The static and dynamical properties of strongly-correlated quantum matter are notoriously difficult to understand. Strong quantum correlations often prohibit intuitive models and the interplay between interactions and kinetic energy gives rise to novel effects, such as quantum magnetism and superconductivity. 
A particular challenge has been the two-dimensional Hubbard model, which is hard to solve on a computer and bears a number of conceptually open questions. However, the simulation of actual materials is (even) more involved and has to go beyond the two-dimensional Hubbard model. Most real materials are not plainly two-dimensional, but possess rather complex lattice structures, which can be approximated as a system of coupled layers.

The simplest realization of a coupled layered material is the bilayer Hubbard model, see Figure 1. In addition to the usual elements of the Hubbard model, namely the tunnel coupling $t$ between adjacent lattice sites and the on-site interaction with energy $U$, it contains the tunnel coupling  $t_{\bot}$ between layers as an independent parameter.

The strength of the tunnel coupling $t_{\bot}$ determines the correlations between the two layers. Hence it plays a pivotal role in determining whether  antiferromagnetic order in each layer is  the dominant configuration or whether even more exotic phases not encountered in pure two-dimensional samples are realized. In case of strong interlayer tunnel coupling this includes a band insulator phase of singlet pairs along the bilayer bonds close to half-filling $n=0.5$ \cite{kancharla2007band,golor2014ground,dos1995magnetism,ruger2014phase}.  In the $U \to \infty$ limit the bilayer Hubbard model maps to the  well-known Heisenberg model with a critical value of 
$t_\perp/t = 1.588$  \cite{sandvik1994order}, below which the system exhibits antiferromagnetic ordering in the layers and above which the system enters the band insulating phase. For decreasing interaction strength, the critical $t_\perp$ that marks the crossover from insulator to band insulator is predicted to increase. Numerical simulations for small systems have revealed that even more exotic phases, such as anti-ferromagnetic metals, could exist \cite{kancharla2007band,hafermann2009metal}. However, the prime experimental challenge to observe these phases is the difficulty of detection. While two-dimensional Hubbard models are now routinely amenable to high-resolution microscopy \cite{Greif2015,Cheuk2016,Cocchi2016}, coupled-layer systems face difficulties for read-out since the layers have to be microscopically close together in order to realize a strong and adjustable coupling. Very recently, techniques to overcome this have been presented \cite{koepsell2020robust,hartke2020measuring} but  the different ground states of the bilayer system have not yet been revealed.

Here, we  realize a bilayer Fermi-Hubbard model using ultracold atoms. We employ both fully spin- and density-resolved imaging techniques with high spatial resolution  to reveal the density and local magnetic correlations. Using tomographic imaging, we are able to directly image both layers separately. Moreover, we measure the staggered magnetic correlation function within and between the layers, thereby revealing the anti-ferromagnetic order. Our results show that the type of magnetic order is highly sensitive to the degree of the interlayer coupling and that we can control  a crossover between two insulators, a planar anti-ferromagnetically ordered Mott insulator and a band insulator of spin-singlets along the bonds between the layers, see Figure 1a.

\begin{figure}
\includegraphics[width=.5\columnwidth]{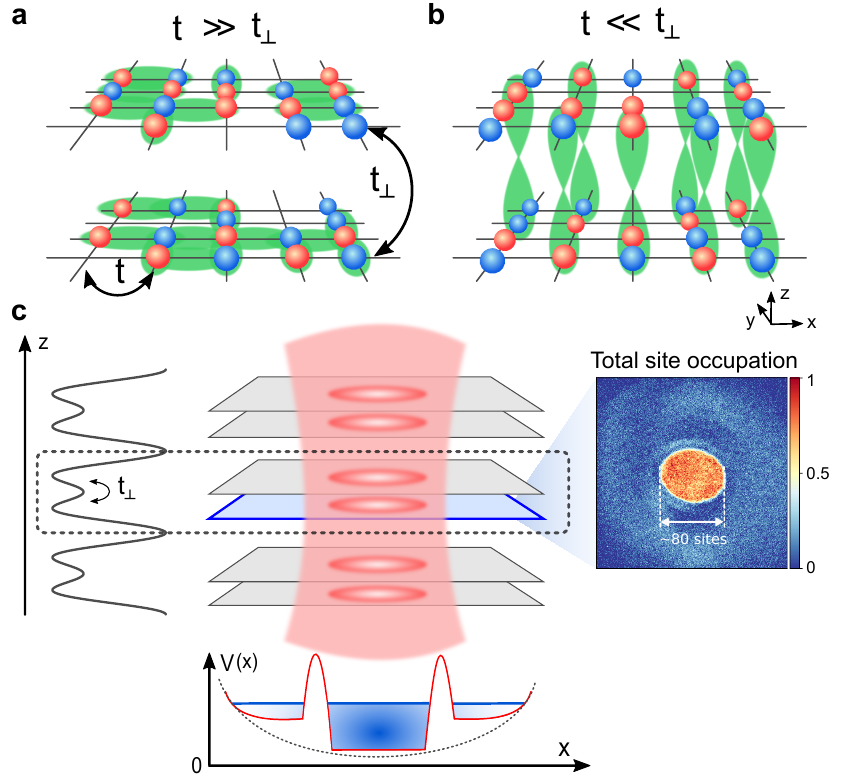}
\caption{Illustration of the bilayer Hubbard model. a) For weak interlayer tunneling ($t_\perp \ll t$) the ground state composes of two weakly-coupled two-dimensional anti-ferromagnetic Mott insulators. The shaded green areas illustrate the spin correlations between spin up and down particles shown as red and blue spheres, respectively.   b) For strong interlayer coupling ($t_\perp \gg t$) the physics is dominated by singlets along the  $z$-axis forming an unconventional band insulator, however, antiferromagnetic correlations in the layer disappear. c) Experimentally, the bilayer system is realized by a bichromatic superlattice in the $z$--direction, trapping atoms in several bilayer sheets. Lateral confinement is provided by optical potentials realizing a flat-bottom trap. Using tomographic imaging (spin-) densities in a single layer are detected. The inset shows the in-situ  density of one layer averaged over 12 realizations.}
	\label{fig1}
\end{figure}

Our experimental setup is an extension of our previous work \cite{Cocchi2016,Wurz2018}. Starting point for the preparation of the bilayer Hubbard model is a two-species band insulator of atoms in the two lowest hyperfine states of $^{40}$K, namely the $\ket{\uparrow}=\ket{F=9/2,m_F=-9/2}$ and $\ket{\downarrow}=\ket{F=9/2,m_F=-7/2}$ states. A 50/50 mixture of these is confined in  a two-dimensional optical lattice in the $xy$-plane with a lattice spacing of $d=532$\,nm. Subsequently, we employ a  bichromatic optical superlattice in the vertical $z$-direction with wavelengths $\lambda_1=532$\,nm and $\lambda_2=1064$\,nm and periods $d_1=1.1\, \mu$m and $d_2=2.2\,\mu$m, respectively, to split the band insulator into two coupled Mott insulators. During the melting of the band insulator, we allow for intra-layer tunneling in the $x$-- and $y$--directions by setting the $xy$ lattice depth to values between $5$ and $7$\,$E_\text{r}$ leading to tunneling amplitudes of $t/h = 290$ to $174$\,Hz. Here, $E_\text{r}=h^2/(8md^2)$ denotes the recoil energy with mass $m$ and Planck's constant $h$. During the splitting procedure, we set the interaction strength to moderately repulsive. Also, we employ a spatial light modulator in order to create a laterally homogeneous trapping potential surrounded by a strong potential barrier that separates off regions of low density which serve as a reservoir for entropy. Our preparation produces a homogeneous bilayer region containing approximately $5600$ sites per layer. For more details see Methods. 



\begin{figure}
\includegraphics[width=.4\columnwidth]{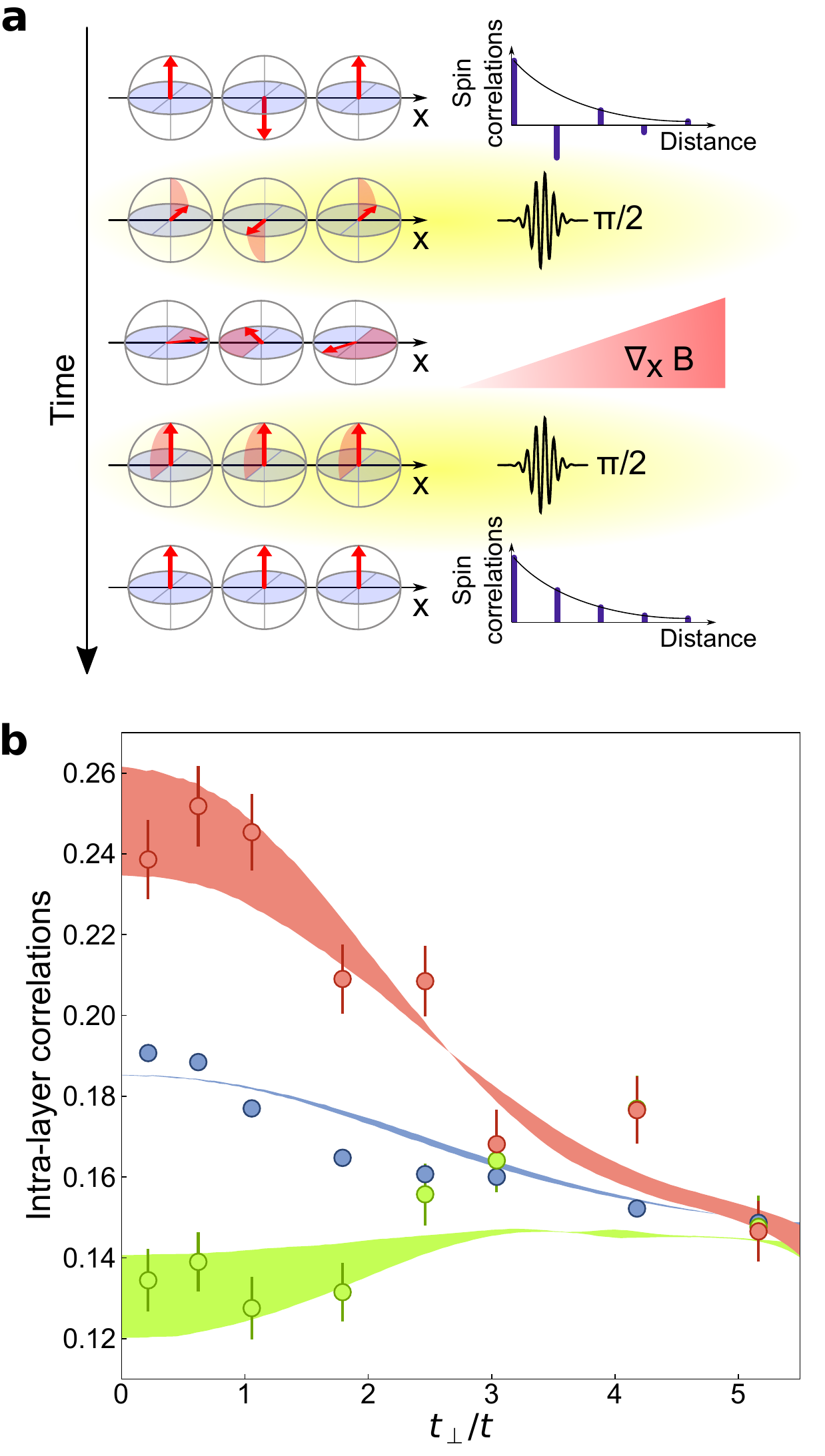}
\caption{Detection of intra-layer correlations. a) Ramsey sequence in a magnetic field gradient for detecting the staggered magnetic structure factor factor $S(\bm{q})$ at wave vector $\bm{q}=(\pi/d,\pi/d)$. b) Staggered structure factor (red) \textit{vs.} interlayer tunneling amplitude $t_\perp$. For comparison, we show also the  uniform structure factor $S(\bm{q}=0)$ (green) and local magnetic moment (blue). The data was taken with $t = 174$\,Hz and $U \simeq 8t$. The shaded areas are the results of DQMC calculations corresponding to the experimental parameters covering the temperature range $k_BT/t = 1.0$ to $1.4$ at filling $n=0.4$. The error bars denote the standard error of the spin correlation results in the center region of the cloud.}
	\label{inlayerCorr}
\end{figure} 

In order to detect the anti-ferromagnetic order within one layer, we measure the staggered magnetic structure factor $S[\bm{q}]$ at wave vector $\bm{q}=(\pi/d,\pi/d)$ in a Ramsey-type experiment, see Figure 2a. To this end, we apply a global $\pi/2$ rotation to all spins, followed by a time evolution in a magnetic field gradient precisely aligned with the diagonal of the $xy$--lattice. The gradient is applied for a time such that spins separated by a distance $\sqrt{2} d$ along the diagonal of the lattice rotate their phase by $2\pi$ relative to each other. Neighbouring sites along the principal lattice axes therefore experience a differential rotation of $\pi$ only. A subsequent $\pi/2$ pulse completes the sequence  and maps the time-evolved spin state into the measurement basis. The density in both spin states is measured by absorption imaging in the same experimental realization. Subsequently, the spin structure factor is measured by an autocorrelation analysis of the difference of the spin-up and spin-down density \cite{Wurz2018}. We combine the Ramsey spin-rotation in each layer of the Hubbard lattice with tomographic resolution in  $z$-direction and hence detect the anti-ferromagnetic correlations in a single layer of the coupled bilayer system.


In Figure 2b, we show the staggered spin structure factor $S[\bm{q}=(\pi/d,\pi/d)]$ for various interlayer tunneling amplitudes $t_{\perp}$. Moreover, we show, for reference purposes, the local magnetic moment $C_0 = \braket{(\hat{S}_i^z)^2}-\braket{\hat{S}_i^z}^2$, which measures the contribution of purely local magnetic correlations without any long-range contribution. Here, $\hat{S}_i^z$ denotes the spin operator on lattice site $i$. The local moment is detected by measuring the density of singly-occupied lattice sites of the two different spin components separately. Finally, we also show the homogeneous magnetic structure factor $S[\bm{q}=(0,0)]$, which is suppressed due to the anti-ferromagnetic ordering. The homogeneous magnetic structure factor is measured using the same autocorrelation analysis as for the staggered spin structure factor, however, without applying the magnetic field gradient prior to detection, see Methods. For systems without any long-range magnetic correlations, all three correlators should be equal to each other. We compare our experimental results to numerical simulations using the Determinant Quantum Monte-Carlo (DQMC) method (see shaded areas in Figure 2b).The simulations describe a system with filling $n=0.4$ to account for imperfections, i.e. holes, in the initial state and local inhomogeneities of the trap potential. All three magnetic correlators agree very well with the experimental data. At the temperatures reached in our experiment we do not expect long-range correlations. This is reflected in the distance of the staggered and uniform structure factor to the local moment being equal, which indicates nearest-neighbour correlations only.
In particular for very large values of $t_\perp$
 we observe, that the homogeneous and staggered structure factors agree within errors with each other, which directly implies that within the layer there are only on-site spin correlations. The intra-layer spin correlation data is particularly sensitive to any imperfection in the detection fidelity of the monolayer tomography. For the data set presented here we ensured that the contributions from neighbouring planes are negligible.

We observe that the anti-ferromagnetic intra-layer correlations disappear for increasing coupling $t_{\perp}$ between the two-dimensional layers. This result is in stark contrast to the full three-dimensional Hubbard model, where anti-ferromagnetic correlations in all directions are enhanced by a higher coordination number than in two dimensions, which, together with reduced quantum fluctuations, leads to a phase transition at finite temperature \cite{scalettar1995magnetism}.  However, for a bilayer system, an increasing $t_{\perp}$ has been theoretically predicted  \cite{scalettar1994magnetic,sandvik1994order,kancharla2007band,golor2014ground,dos1995magnetism,ruger2014phase}  to drive the formation of singlets across the bonds between the two layers at the expense of reducing magnetic correlations within the layers, as we shall demonstrate experimentally next.

\begin{figure}
\centering
\includegraphics[width=.4\columnwidth]{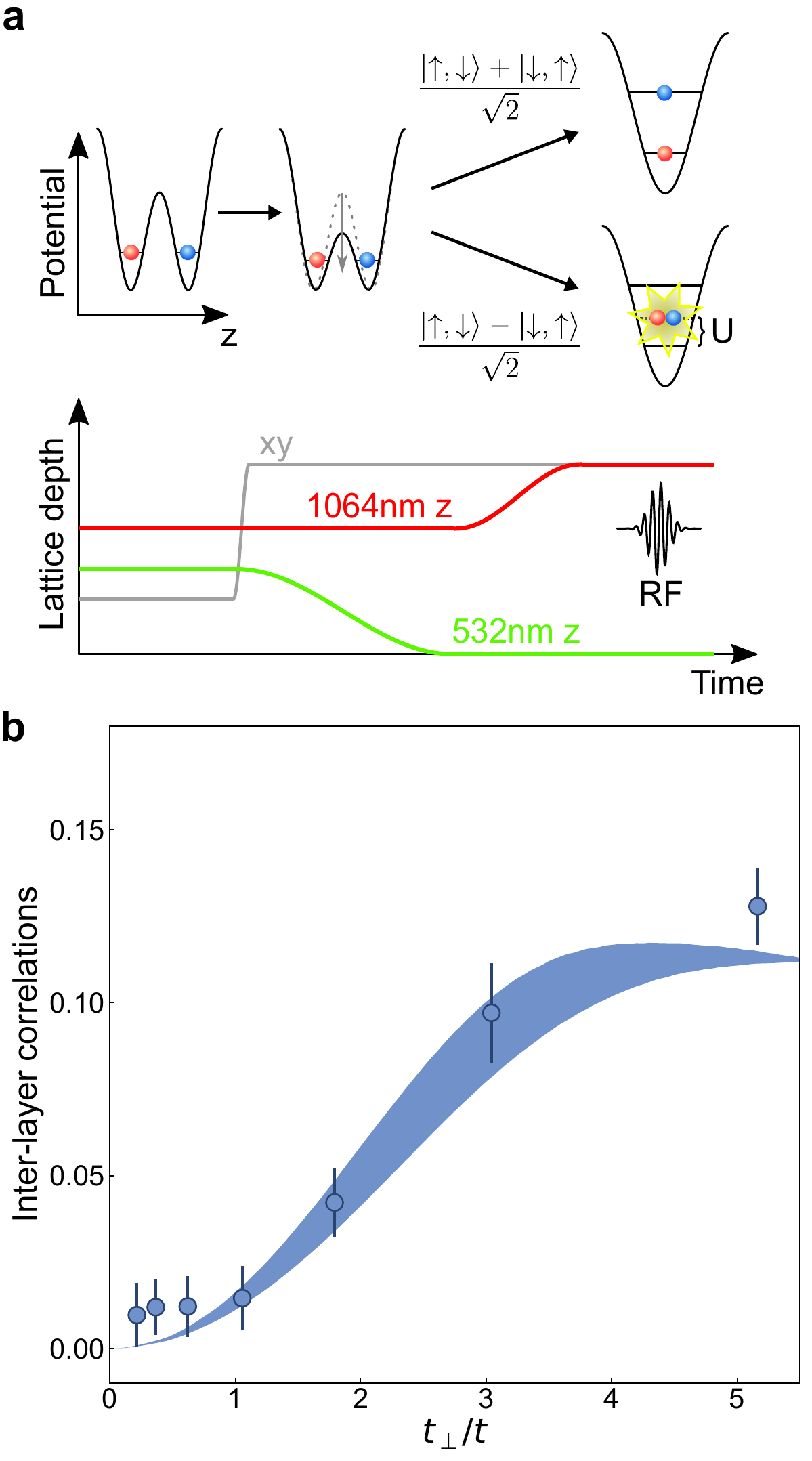}
\caption{Magnetic correlations between the layers. a) A double well along the $z$--direction is merged by a suitable ramp of the intensities of the optical superlattice. Only if the double-well has been in a spin-singlet configuration both atoms end up in the vibrational ground state of the merged lattice, which is detected by radio-frequency spectroscopy. b) Staggered spin correlator between the coupled layers of the bilayer system. The data was taken with the same Hubbard parameters as for Figure 2b. The error bars denote the standard deviation of the interlayer correlations for different realizations. The shaded area shows the DQMC calculations for a temperature range from $k_B T = 1.0t$ to $1.4t$. }
\label{interlayerCorr}
\end{figure} 

We measure the interlayer magnetic correlations using the technique shown in Figure 3a \cite{greif2013short}. After having created the bilayer system, we rapidly freeze the motion in the $xy$--layers and thereby effectively create an array of separated double wells along the $z$-axis. Each double well can be occupied by up to four fermions. At half-filling, the large majority of double wells will be  in either a spin-singlet state $(\ket{\uparrow, \downarrow} -\ket{\downarrow, \uparrow})/\sqrt{2}$ or a triplet state $\left \{\ket{\uparrow, \uparrow}, (\ket{\uparrow, \downarrow} +\ket{\downarrow,\uparrow})/\sqrt{2}, \ket{\downarrow, \downarrow} \right\}$, which we will discuss exemplary, however, the conclusion from the following argument is valid for any occupation. By adiabatically reducing the potential barrier between the two wells, the separated atoms will merge into one single well. To maintain the overall anti-symmetry of the two-fermion wave function, only the anti-symmetric spin-singlet state  merges into the vibrational ground states. In contrast, when merging a spin-triplet state, one atom ends up in a higher vibrational level of the lattice. We distinguish both outcomes  and determine the probability of a doubly-occupied vibrational ground state by performing radiofrequency spectroscopy, which resolves the on-site interaction shift $U$ \cite{greif2013short}, combined with in-situ imaging. After subtracting the average double occupancy in both layers measured without merging, the doubles density is proportional to the probability $p_\text{dimer}$ of anti-ferromagnetic spin-singlets along a bond between the  coupled layers. This probability is converted into a staggered spin correlator $C_z = -(\braket{\hat{S}_{i1}^z\hat{S}_{i2}^z}-\braket{\hat{S}_{i1}^z}\braket{\hat{S}_{i2}^z})=  p_\text{dimer}/4$. The factor of $1/4$ results from the consideration that if each bond is occupied by a singlet state, the spin correlator between the layers should match the double-well expectation value of $C_z=1/4$.

Figure~3b shows the measured inter-layer correlations as a function of $t_\perp$. We observe that increasing $t_\perp$ enhances the inter-layer correlations, which is a key feature of the band insulator phase. Furthermore, they show the opposite behaviour as compared to the intra-layer correlations shown in Figure 2b. Therefore we conclude that  by tuning the interlayer coupling, we observe the crossover from the antiferromagnetic Mott insulator to the band insulator.


Finally, in Figure 4 we show how the crossover depends on the interaction strength $U$. To this end, we analyze the ratio of total intra-layer correlations $C_{xy}$ and inter-layer correlations $C_z$ by $R=\frac{C_{xy}}{C_{xy}+C_z}$. The total intra-layer correlations are defined as $C_{xy}=2(S[\bm{q}=(\pi/d,\pi/d)]-C_0)$, where we subtract the local moment from the staggered spin structure factor in order to take into account only non-local spin correlations and multiply by a factor of two to account for the two layers. The ratio $R$ ranges from one for purely intra-layer magnetic correlations to zero for purely inter-layer correlations. The results show a crossover between the limits. We have interpolated the data to extract the value of $R=0.5$, which we show as crosses in the figure. From this we conclude that the crossover  occurs at $t_\perp/t \simeq 2.5$ with  only a weak dependence on the interaction strength. Results from numerical simulations \cite{bouadim2008magnetic} show a similar behaviour.

We further investigate the insulating character of the bilayer system for varying $t_\perp$. By applying an in-plane magnetic field gradient of strength $\left|\nabla B_z\right| = 24.8\ \mathrm{G/m}$, and according to the local-density approximation, we extract the filling $n$ of a single layer for different chemical potentials. With this we can calculate the isothermal compressibility $\kappa = \partial n/\partial \mu$ close to half-filling. Our results show the increasingly insulating nature when approaching the band insulating state at high $t_\perp$ and agree with DQMC calculations.

Our work shows that strongly-interacting lattice systems composed of coupled layers show qualitatively different regimes as compared to the two-dimensional Hubbard model. The insights and methodology developed here open the route towards quantum simulations of real materials and, with modifications to the intra-layer lattices, of bilayer Haldane models or stacked graphene-like materials.

\begin{figure}
	\centering
	\includegraphics[width=.3\columnwidth]{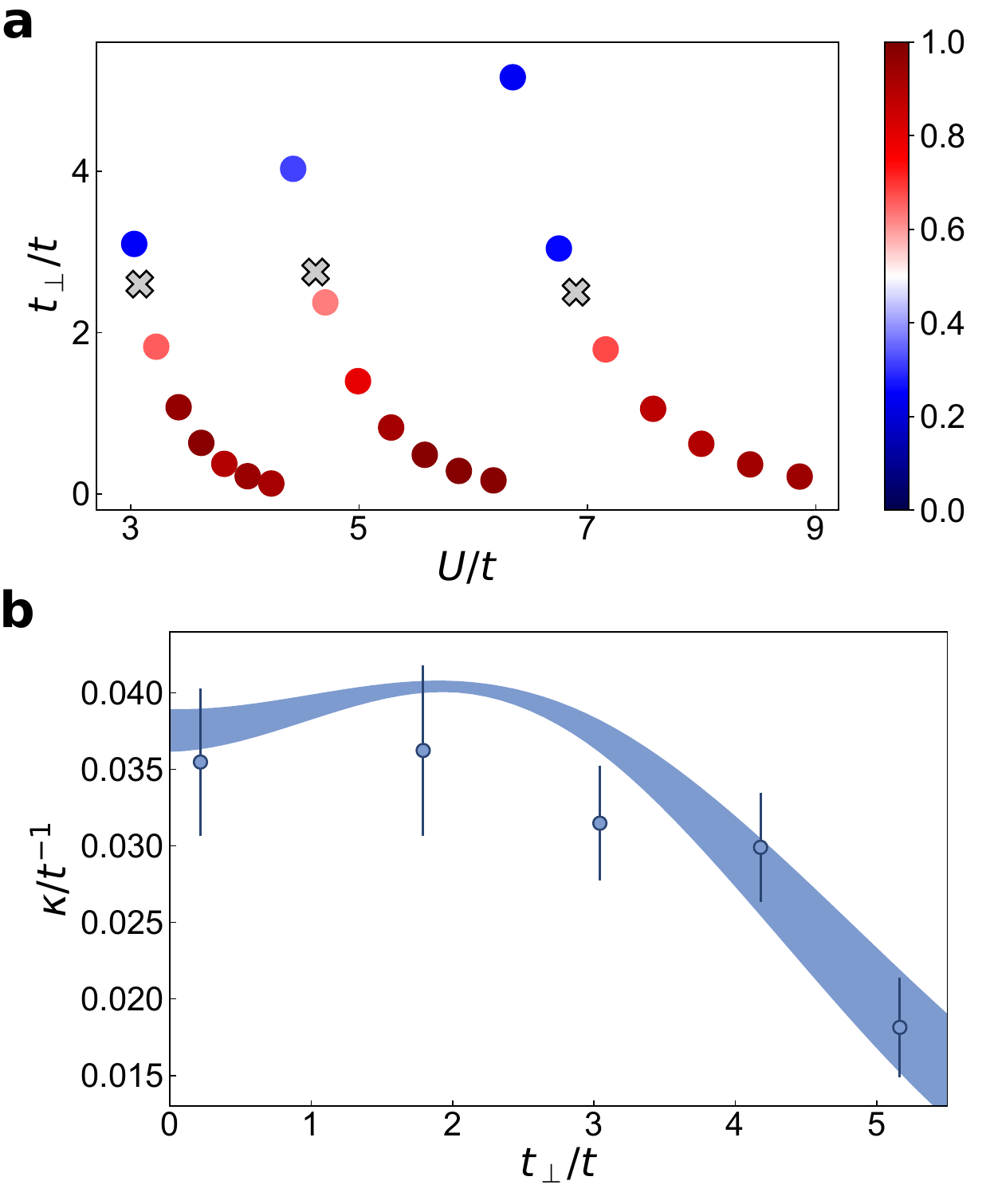}
	\caption{Crossover from the anti-ferromagnetic Mott insulator to a band insulator of singlets. a) We plot the ratio $R$ of  intra-layer to the sum of intra- and inter-layer magnetic correlations as a function of both $t_\perp/t$ and $U/t$. The crosses show the interpolated values for $R=0.5$. The relative standard error on the ratio $R$ is approximately $17\%$. b) Measurement of the compressibility as a function of the interlayer tunneling amplitude $t_\perp$ at $U/t\approx 8$. The error bars denote the standard deviation. The shaded region shows the DQMC calculations for a temperature range of $k_B T= 1.0t$ to $1.4t$. }
	\label{interlayerCorr}
\end{figure}

This work has been supported by BCGS, the Alexander-von-Humboldt Stiftung, DFG (SFB/TR 185 project B4),  Cluster of Excellence Matter and Light for Quantum Computing (ML4Q) EXC 2004/1 - 390534769 and Stiftung der deutschen Wirtschaft.

\vspace{0.5 cm}
\noindent {\bf Data Availability}\\
The data presented in the figures is available on https://osf.io/u9wj6. More detailed data and information of this study is available from the corresponding author upon request.
\vspace{0.5 cm}

\noindent {\bf Code availability}\\
The DQMC theory is simulated using the QUantum Electron Simulation Toolbox (QUEST) Fortran 90/95 package, version 1.44, from https://code.google.com/archive/p/quest-qmc/.

\vspace{0.5 cm}
\noindent {\bf Competing interest}\\
The authors declare that they have no competing interests.
\vspace{0.5 cm}

\vspace{0.5 cm}
\noindent {\bf Author contributions}\\
The experiment was perceived by M.G. N.W., C.C. and M.K., data taking was performed by M.G., N.W. and C.C. with contributions by J.S., data analysis was primarily performed by M.G. and N.W, numerical simulations were performed by C.C. and N.W., the results were discussed and interpreted by all coauthors, and the manuscript was written by M.K. with contributions from all coauthors.
\vspace{0.5 cm}

\vspace{0.5 cm}
\noindent {\bf Corresponding authors}\\
The corresponding author is Michael K\"ohl (email: michael.koehl@uni-bonn.de).


\vspace{0.5 cm}

\begin{thebibliography}{10}
	\expandafter\ifx\csname url\endcsname\relax
	\def\url#1{\texttt{#1}}\fi
	\expandafter\ifx\csname urlprefix\endcsname\relax\def\urlprefix{URL }\fi
	\providecommand{\bibinfo}[2]{#2}
	\providecommand{\eprint}[2][]{\url{#2}}
	
	\bibitem{Greif2015}
	\bibinfo{author}{Greif, D.} \emph{et~al.}
	\newblock \bibinfo{title}{Site-resolved imaging of a fermionic Mott
		insulator}.
	\newblock \emph{\bibinfo{journal}{Science}} \textbf{\bibinfo{volume}{351}},
	\bibinfo{pages}{953--957} (\bibinfo{year}{2016}).
	
	\bibitem{Cheuk2016}
	\bibinfo{author}{Cheuk, L.~W.} \emph{et~al.}
	\newblock \bibinfo{title}{Observation of 2d fermionic {M}ott insulators of
		$^{40}\mathrm{K}$ with single-site resolution}.
	\newblock \emph{\bibinfo{journal}{Phys. Rev. Lett.}}
	\textbf{\bibinfo{volume}{116}}, \bibinfo{pages}{235301}
	(\bibinfo{year}{2016}).
	\newblock
	\urlprefix\url{http://link.aps.org/doi/10.1103/PhysRevLett.116.235301}.
	
	\bibitem{Cocchi2016}
	\bibinfo{author}{Cocchi, E.} \emph{et~al.}
	\newblock \bibinfo{title}{Equation of state of the two-dimensional {H}ubbard
		model}.
	\newblock \emph{\bibinfo{journal}{Phys. Rev. Lett.}}
	\textbf{\bibinfo{volume}{116}}, \bibinfo{pages}{175301}
	(\bibinfo{year}{2016}).
	\newblock
	\urlprefix\url{http://link.aps.org/doi/10.1103/PhysRevLett.116.175301}.
	
	\bibitem{Cheuk2016b}
	\bibinfo{author}{Cheuk, L.~W.} \emph{et~al.}
	\newblock \bibinfo{title}{Observation of spatial charge and spin correlations
		in the {2D} {F}ermi-{H}ubbard model}.
	\newblock \emph{\bibinfo{journal}{Science}} \textbf{\bibinfo{volume}{353}},
	\bibinfo{pages}{1260--1264} (\bibinfo{year}{2016}).
	\newblock \urlprefix\url{http://science.sciencemag.org/content/353/6305/1260}.
	
	\bibitem{Parsons2016}
	\bibinfo{author}{Parsons, M.~F.} \emph{et~al.}
	\newblock \bibinfo{title}{Site-resolved measurement of the spin-correlation
		function in the {F}ermi-{H}ubbard model}.
	\newblock \emph{\bibinfo{journal}{Science}} \textbf{\bibinfo{volume}{353}},
	\bibinfo{pages}{1253--1256} (\bibinfo{year}{2016}).
	\newblock \urlprefix\url{http://science.sciencemag.org/content/353/6305/1253}.
	
	\bibitem{Drewes2017}
	\bibinfo{author}{Drewes, J.~H.} \emph{et~al.}
	\newblock \bibinfo{title}{Antiferromagnetic correlations in two-dimensional
		fermionic {M}ott-insulating and metallic phases}.
	\newblock \emph{\bibinfo{journal}{Phys. Rev. Lett.}}
	\textbf{\bibinfo{volume}{118}}, \bibinfo{pages}{170401}
	(\bibinfo{year}{2017}).
	\newblock
	\urlprefix\url{https://link.aps.org/doi/10.1103/PhysRevLett.118.170401}.
	
	\bibitem{mazurenko2017cold}
	\bibinfo{author}{Mazurenko, A.} \emph{et~al.}
	\newblock \bibinfo{title}{A cold-atom Fermi--Hubbard antiferromagnet}.
	\newblock \emph{\bibinfo{journal}{Nature}} \textbf{\bibinfo{volume}{545}},
	\bibinfo{pages}{462--466} (\bibinfo{year}{2017}).
	
	\bibitem{mitra2018quantum}
	\bibinfo{author}{Mitra, D.} \emph{et~al.}
	\newblock \bibinfo{title}{Quantum gas microscopy of an attractive
		Fermi--Hubbard system}.
	\newblock \emph{\bibinfo{journal}{Nature Physics}}
	\textbf{\bibinfo{volume}{14}}, \bibinfo{pages}{173--177}
	(\bibinfo{year}{2018}).
	
	\bibitem{scalettar1994magnetic}
	\bibinfo{author}{Scalettar, R.~T.}, \bibinfo{author}{Cannon, J.~W.},
	\bibinfo{author}{Scalapino, D.~J.} \& \bibinfo{author}{Sugar, R.~L.}
	\newblock \bibinfo{title}{Magnetic and pairing correlations in coupled Hubbard
		planes}.
	\newblock \emph{\bibinfo{journal}{Physical Review B}}
	\textbf{\bibinfo{volume}{50}}, \bibinfo{pages}{13419} (\bibinfo{year}{1994}).
	
	\bibitem{maier2011pair}
	\bibinfo{author}{Maier, T.~A.} \& \bibinfo{author}{Scalapino, D.}
	\newblock \bibinfo{title}{Pair structure and the pairing interaction in a
		bilayer Hubbard model for unconventional superconductivity}.
	\newblock \emph{\bibinfo{journal}{Physical Review B}}
	\textbf{\bibinfo{volume}{84}}, \bibinfo{pages}{180513}
	(\bibinfo{year}{2011}).
	
	\bibitem{kancharla2007band}
	\bibinfo{author}{Kancharla, S.~S.} \& \bibinfo{author}{Okamoto, S.}
	\newblock \bibinfo{title}{Band insulator to Mott insulator transition in a
		bilayer Hubbard model}.
	\newblock \emph{\bibinfo{journal}{Physical Review B}}
	\textbf{\bibinfo{volume}{75}}, \bibinfo{pages}{193103}
	(\bibinfo{year}{2007}).
	
	\bibitem{golor2014ground}
	\bibinfo{author}{Golor, M.}, \bibinfo{author}{Reckling, T.},
	\bibinfo{author}{Classen, L.}, \bibinfo{author}{Scherer, M.~M.} \&
	\bibinfo{author}{Wessel, S.}
	\newblock \bibinfo{title}{Ground-state phase diagram of the half-filled bilayer
		Hubbard model}.
	\newblock \emph{\bibinfo{journal}{Physical Review B}}
	\textbf{\bibinfo{volume}{90}}, \bibinfo{pages}{195131}
	(\bibinfo{year}{2014}).
	
	\bibitem{dos1995magnetism}
	\bibinfo{author}{Dos~Santos, R.~R.}
	\newblock \bibinfo{title}{Magnetism and pairing in Hubbard bilayers}.
	\newblock \emph{\bibinfo{journal}{Physical Review B}}
	\textbf{\bibinfo{volume}{51}}, \bibinfo{pages}{15540} (\bibinfo{year}{1995}).
	
	\bibitem{ruger2014phase}
	\bibinfo{author}{R{\"u}ger, R.}, \bibinfo{author}{Tocchio, L.~F.},
	\bibinfo{author}{Valent{\'\i}, R.} \& \bibinfo{author}{Gros, C.}
	\newblock \bibinfo{title}{The phase diagram of the square lattice bilayer
		Hubbard model: a variational Monte Carlo study}.
	\newblock \emph{\bibinfo{journal}{New Journal of Physics}}
	\textbf{\bibinfo{volume}{16}}, \bibinfo{pages}{033010}
	(\bibinfo{year}{2014}).
	
	\bibitem{sandvik1994order}
	\bibinfo{author}{Sandvik, A.} \& \bibinfo{author}{Scalapino, D.}
	\newblock \bibinfo{title}{Order-disorder transition in a two-layer quantum
		antiferromagnet}.
	\newblock \emph{\bibinfo{journal}{Physical review letters}}
	\textbf{\bibinfo{volume}{72}}, \bibinfo{pages}{2777} (\bibinfo{year}{1994}).
	
	\bibitem{hafermann2009metal}
	\bibinfo{author}{Hafermann, H.}, \bibinfo{author}{Katsnelson, M.} \&
	\bibinfo{author}{Lichtenstein, A.}
	\newblock \bibinfo{title}{Metal-insulator transition by suppression of spin
		fluctuations}.
	\newblock \emph{\bibinfo{journal}{EPL (Europhysics Letters)}}
	\textbf{\bibinfo{volume}{85}}, \bibinfo{pages}{37006} (\bibinfo{year}{2009}).
	
	\bibitem{koepsell2020robust}
	\bibinfo{author}{Koepsell, J.} \emph{et~al.}
	\newblock \bibinfo{title}{Robust bilayer charge-pumping for spin-and
		density-resolved quantum gas microscopy}.
	\newblock \emph{\bibinfo{journal}{arXiv preprint arXiv:2002.07577}}
	(\bibinfo{year}{2020}).
	
	\bibitem{hartke2020measuring}
	\bibinfo{author}{Hartke, T.}, \bibinfo{author}{Oreg, B.}, \bibinfo{author}{Jia,
		N.} \& \bibinfo{author}{Zwierlein, M.}
	\newblock \bibinfo{title}{Measuring total density correlations in a
		Fermi-Hubbard gas via bilayer microscopy} (\bibinfo{year}{2020}).
	\newblock \eprint{2003.11669}.
	
	\bibitem{Wurz2018}
	\bibinfo{author}{Wurz, N.} \emph{et~al.}
	\newblock \bibinfo{title}{Coherent manipulation of spin correlations in the
		{H}ubbard model}.
	\newblock \emph{\bibinfo{journal}{Phys. Rev. A}} \textbf{\bibinfo{volume}{97}},
	\bibinfo{pages}{051602} (\bibinfo{year}{2018}).
	\newblock \urlprefix\url{https://link.aps.org/doi/10.1103/PhysRevA.97.051602}.
	
	\bibitem{scalettar1995magnetism}
	\bibinfo{author}{Scalettar, R.~T.}
	\newblock \bibinfo{title}{Magnetism and spin liquid behavior in a two layer
		Hubbard model}.
	\newblock \emph{\bibinfo{journal}{Journal of low temperature physics}}
	\textbf{\bibinfo{volume}{99}}, \bibinfo{pages}{499--504}
	(\bibinfo{year}{1995}).
	
	\bibitem{greif2013short}
	\bibinfo{author}{Greif, D.}, \bibinfo{author}{Uehlinger, T.},
	\bibinfo{author}{Jotzu, G.}, \bibinfo{author}{Tarruell, L.} \&
	\bibinfo{author}{Esslinger, T.}
	\newblock \bibinfo{title}{Short-range quantum magnetism of ultracold fermions
		in an optical lattice}.
	\newblock \emph{\bibinfo{journal}{Science}} \textbf{\bibinfo{volume}{340}},
	\bibinfo{pages}{1307--1310} (\bibinfo{year}{2013}).
	
	\bibitem{bouadim2008magnetic}
	\bibinfo{author}{Bouadim, K.}, \bibinfo{author}{Batrouni, G.~G.},
	\bibinfo{author}{H{\'e}bert, F.} \& \bibinfo{author}{Scalettar, R.}
	\newblock \bibinfo{title}{Magnetic and transport properties of a coupled
		hubbard bilayer with electron and hole doping}.
	\newblock \emph{\bibinfo{journal}{Physical Review B}}
	\textbf{\bibinfo{volume}{77}}, \bibinfo{pages}{144527}
	(\bibinfo{year}{2008}).
	
	\bibitem{Varney2009}
	\bibinfo{author}{Varney, C.~N.} \emph{et~al.}
	\newblock \bibinfo{title}{Quantum Monte Carlo study of the two-dimensional
		fermion {H}ubbard model}.
	\newblock \emph{\bibinfo{journal}{Phys. Rev. B}} \textbf{\bibinfo{volume}{80}},
	\bibinfo{pages}{075116} (\bibinfo{year}{2009}).
	\newblock \urlprefix\url{http://link.aps.org/doi/10.1103/PhysRevB.80.075116}.
	
\end{thebibliography}

\section{Methods}

\subsection{Bilayer Hubbard Hamiltonian}

The Hamiltonian describing our system contains the tunnelling amplitude $t$ between neighbouring sites $i$ and $j$ of the same layer $m$, as well as the tunnelling amplitude between the two layers $t_\perp$. Here, $\hat{c}_{im,\sigma}^\dagger$ denotes the creation operator at lattice site $i$ in layer $m$ with spin $\sigma$. Doubly occupied sites experience a shift in energy $U$.  The chemical potential $\mu$ fixes the average filling $\left< \hat{n}_{im,\sigma}\right>$,  where $n_{im,\sigma}$ describes the density at lattice site $im$ in spin state $\sigma$. 

\begin{align*}
\hat{H} = -t \sum_{\left<ij\right>m,\sigma} \hat{c}_{im,\sigma}^\dagger \hat{c}_{jm,\sigma} -t_\perp\sum_{i,\sigma} \left(\hat{c}_{i1,\sigma}^\dagger \hat{c}_{i2,\sigma} + h.c.\right) + U \sum_{im} \hat{n}_{im,\uparrow}\hat{n}_{im,\downarrow}  -\mu \sum_{im,\sigma} \hat{n}_{im,\sigma}
\end{align*}

\subsection{Loading the bilayer}

Initially, we prepare a band insulator of two spin states encoded in the two lowest hyperfine states of $^{40}$K $\ket{\uparrow}=\ket{F=9/2,m_F=-9/2}$ and $\ket{\downarrow}=\ket{F=9/2,m_F=-7/2}$ at attractive interactions of $U/t= -1.7$. This ensures a high occupation  of $n = 0.95$ per lattice site. For this preparation, the  phase between the superlattices has been adjusted such that only every second layer of the lattice is populated and tunneling to neighbouring layers is suppressed. Additionally, during the lattice loading, we ramp up an optical potential created by the spatial light modulator at the outer regions of the atomic cloud to increase the density at the center. Subsequently, we  freeze the density distribution of the band insulator by quickly increasing the intra-layer ($xy$--) lattice depth. In order to prepare a repulsively interacting gas, we apply a radiofrequency pulse on the $\ket{F=9/2,m_F=-7/2} \to \ket{F=9/2,m_F=-5/2}$ transition, ramp the magnetic field below the Feshbach resonance of the $\ket{F=9/2,m_F=-9/2}/\ket{F=9/2,m_F=-7/2}$ states and apply a second radiofrequency pulse to convert $\ket{F=9/2,m_F=-5/2} \to \ket{F=9/2,m_F=-7/2}$. The filling reduces to $n = 0.9$ upon transferring the band insulator from attractive to repulsive interactions. Subsequently, we shift the superlattice phase closer to the symmetry point and  increase the power of the short-wavelength $z$-lattice, which slowly splits the band insulator into a bilayer lattice close to half filling. The choice of the final superlattice phase allows to adjust and correct any potential energy offset between the two layers, for example gravitational sag. 

\subsection{Tomographic in-situ imaging of a single layer} 
After preparing the atoms in the stack of bilayer systems, we freeze their motion by ramping up the horizontal lattice depth within $1\,$ms to a value of $60\,E_{rec}$ and, simultaneously, the short-wavelength $z$--lattice to $110\,E_{rec}$. In order to detect a single two-dimensional layer, a strong vertical magnetic field gradient in $z$--direction is applied, allowing to resolve the magnetic field sensitive hyperfine transition frequencies of the layers. Using radio frequency (RF) tomography the atoms of one layer are then transferred to another internal state for detection. Subsequently, we implement a spin-/density-resolved detection protocol for the measurement of the intra-/ inter-layer spin correlations, respectively. For the inter-layer correlations we need to distinguish singly- and doubly-occupied sites, which we achieve with another RF transfer that resolves the difference in on-site interaction between initial and final state of 1.8kHz. For the intra-layer correlations we employ a spin-resolved measurement, making use of the spin-changing collision between $\ket{F = 9/2, m_F = -9/2}$ and $\ket{F = 9/2, m_F = -3/2}$ to remove doubly-occupied sites. Finally, absorption images of singles and doubles or spin-up and spin-down singles are taken \cite{Cocchi2016,Drewes2017}.

\subsection{Calibration of Hubbard parameters}
We characterize the Hubbard parameter $U$ in the final lattice configuration by radiofrequency spectroscopy of the energy shift caused by on-site interactions. We observe a decrease of $U$ of $20\%$ going from low to high $t_\perp$ due to the decreased compression of the Wannier wave function. Additionally, we calibrate the tunneling amplitude $t_\perp$ and the energy offset $\Delta$ of the double well using a spin-polarized atomic cloud in a deep $xy$-lattice, forming separated double wells in the $z$--direction. Initially, we populate only one well of the double-well configuration before  quickly reducing the intensity of the short-wavelength $z$-lattice in order to induce Rabi tunnel oscillations.



\subsection{Structure factor measurement}
In our experiment we measure the two-dimensional spin structure factor at wave vector $\bm{q}$
\begin{align*}
S(\bm{q}) = \frac{1}{N}\sum_{i,j} e^{-i \bm{q}\cdot \bm{r}_{ij}}C^z_{ij}\, 
\end{align*}
within each layer of the bilayer system \cite{Drewes2017,Wurz2018}. Here, $\bm{r}_{ij}=\bm{r}_{j}-\bm{r}_{i}$ is the distance between lattice sites $i$ and $j$, $N$ is the number of lattice sites, and $C^z_{ij} = \braket{ \hat{S}_i^z \hat{S}_j^z}-\braket{ \hat{S}_i^z} \braket{\hat{S}_j^z}$ denotes the spin correlator between sites $i$ and $j$. The operator $\hat{S}^z_j = (\hat{n}_{j,\uparrow}-\hat{n}_{j,\downarrow})/2$ defines the on-site magnetization.

The uniform structure factor $S[\bm{q}=(0,0)]$ is measured by the autocorrelation analysis of the difference of two absorption images of the spin-up and spin-down densities taken in one realization of the experiment. The staggered magnetic structure factor at wave vector $q=(\pi/d,\pi/d)$ is measured by using the spin-spiral imprinting technique discussed in the main text. In contrast, the local moment is directly inferred from the singles densities, since only singly-occupied sites add to the local magnetization, hence
the local moment is $C_{00} = (\braket{\hat{s}_\uparrow}+\braket{\hat{s}_\downarrow})/4$. The staggered and uniform structure factor will approach this value once the off-site correlators go to zero in an uncorrelated system, for example at high temperature.

\subsection{DQMC simulation}

The DQMC simulations are performed using the Quantum Electron Simulation Toolbox (QUEST) Fortran package \cite{Varney2009}. Simulations are performed for a homogeneous lattice with $8 \times 8 \times 2$ sites with $2000$ warm-up sweeps and $200000$ measurement sweeps, and the number of imaginary time slices is set to $25$. For the numerical data shown in the manuscript, the inter-layer tunnelling is varied from $t_\perp /t =  0$ to $4.5$ and on-site repulsion is varied from $U/t = 2$ to $8$. Small doping is introduced by varying the chemical potential over the range of $\mu/t = 0$ to $-2.5$, which corresponds to approximately a filling ranging from $n=0.5$ to $0.4$. The magnetic structure factor is obtained by a finite Fourier transform of the spatial spin correlators.

\end{document}